\documentclass[aps,prc,superscriptaddress,twoside,twocolumn,nofootinbib,10pt,%
showpacs,floatfix]{revtex4-1}

\usepackage{amsmath,amssymb}
\usepackage{graphicx,bm}
\usepackage{slashed}
\usepackage{epstopdf}
\usepackage{ulem} 
\usepackage[usenames]{color}
\usepackage{float}

\renewcommand\sout{\bgroup \color{red} \ULdepth=-.5ex \ULset}

\begin{document}
\title{$P_c(4380)$ in a constituent quark model}

\author{Woosung Park}\email{diracdelta@hanmail.net}\affiliation{Department of Physics and Institute of Physics and Applied Physics, Yonsei University, Seoul 03722, Korea}
\author{Aaron Park}
\email{lacid0220@naver.com}\affiliation{Department of Physics and
Institute of Physics and Applied Physics, Yonsei University, Seoul
03722, Korea}
\author{Sungtae
Cho}\email{sungtae.cho@kangwon.ac.kr}\affiliation{Division of
Science Education, Kangwon National University, Chuncheon 24341,
Korea}
\author{Su Houng Lee}\email{Corresponding author; suhoung@yonsei.ac.kr}\affiliation{Department of Physics and Institute of Physics and Applied Physics, Yonsei University, Seoul 03722, Korea}
\date{\today}
\begin{abstract}
The constituent quark model with color-spin hyperfine potential is
used to investigate the property of a compact pentaquark
configuration with $J^p$=$3/2^-$ and isospin=1/2, which is the
most likely quantum number of one of the recently observed exotic
baryon states at LHCb. Starting from the characterization of the
isospin, color, and spin states for the pentaquark configuration,
we construct the total wave function composed of the spatial wave
function, which we take to be symmetric and in S-wave, and the
four orthogonal isospin $\otimes$ color $\otimes$ spin states that
satisfy the Pauli principle. We then use the variational method to
find a compact stable configuration. While there are compact
configurations where the hyperfine potential is more attractive
than the sum of $p$ and $J/\psi$ hyperfine potentials, we find that the
ground state is the  isolated $p$ and $J/\psi$ state. Furthermore,
the mass of the excited state lies far above the observed
pentaquark state leading us to conclude that the observed states
can not be a compact multiquark configuration with $J^p$=$3/2^-$.
\end{abstract}


\maketitle

\section{Introduction}
After the introduction of the quark model for the baryon and
meson~\cite{GellMann:1964nj} and the
color quantum number for quarks \cite{Han:1965pf}, model
calculations for hadrons natural led  to the possible
existence of mutiquark hadrons beyond the normal hadrons
\cite{Jaffe:1976ig,Jaffe:1976ih}. Indeed, recent experimental
findings point to the possible existence of such configurations;
these are the $XYZ$ states with the $X(3872)$ being the first of
these states observed by the Belle
collaboration~\cite{Choi:2003ue}. The $XYZ$ states could be either compact tetraquark states composed of two quarks
and two antiquarks or  molecular states with their
masses close to the relevant two meson thresholds.

Molecular configurations involving heavy mesons were
first discussed in Ref.~\cite{Tornqvist:1993ng} where deuteronlike meson-meson bound states were found to exist when a long range pion exchange potential was included  with additional short range attraction  depending on the mass of the meson.
The possible bound states included a $D{\bar{D}}^*$ state  in the isopin 0 and $J^{PC}=1^{++}$ channel, which is the quantum numbers of the $X(3872)$.  After the  experimental observation of $X(3872)$, attempts to explain the state in terms of molecular configuration with important contribution coming from the pion exchange potentials still continues to this date~
\cite{Thomas:2008ja,Lee:2009hy,Wang:2013kva,Baru:2015nea,Hosaka:2016pey}.

Numerous efforts have been made to explain the mass of the charmonium-like state using various other approaches. In a
non-relativistic quark model that includes
a confining interaction and a short range
spin-dependent interaction through the one gluon exchange as well
as an effective pion-induced interaction, it was argued that the
$X(3872)$ can be a $D{\bar{D}}^*$ hadronic resonance with important
admixtures of $\rho J/\psi$ and $\omega J/\psi$ states~\cite{Swanson:2003tb}. In Ref.~\cite{Wong:2003xk}, the
$X(3872)$ was considered as a weakly bound molecular state found
in the combination of $\{D,D^*\}$ with $\{\bar{D},{\bar{D}}^*\}$ states
based on a quark based non-relativistic four-body
Hamiltonian with a pairwise interaction.

There are also models that find $X(3872)$ to be a
tetraquark system.  These include methods based on a diquark-antidiquark
model~\cite{Maiani:2004vq,Ebert:2005nc}, the QCD sum
rule~\cite{Chen:2010ze}, and a simple quark model with
chromomagnetic
interactions~\cite{Hogaasen:2005jv,Cui:2006mp,Buccella:2006fn}.
 In a lattice QCD
calculation~\cite{Padmanath:2015era}, it was
shown that a candidate for $X(3872)$ with $I=0$
could only be found if both the $\bar{c}c$ and
$\bar{D}{\bar{D}}^*$ interpolators are included, while no signal
was found if diquark-antidiquark and $\bar{D}{\bar{D}}^*$ are used
without a $\bar{c}c$ component.

Recently, the observation of hidden-charm pentaquark states by the
LHCb collaboration~\cite{Aaij:2015tga}, has triggered another wave
of works among many researchers. The $J/\psi p$ invariant mass
spectrum of ${\Lambda}_b$ $\to$ $J/\psi K^- p$ revealed
hidden-charm pentaquark states, for which the preferred quantum
numbers are $J^p$=$3/2^-$ for $P_c(4380)$ and $J^p$=$5/2^+$ for
$P_c(4450)$. In fact, even before the discovery was made, possible
hidden-charm molecular baryons composed of anti-charmed meson and
charmed baryon, such as the of ${\Sigma}_c\bar{D}^*$ states with
$I(J^p)$=$\frac{1}{2}({\frac{1}{2}}^-)$,
$\frac{1}{2}({\frac{3}{2}}^-)$, $\frac{3}{2}({\frac{1}{2}}^-)$,
$\frac{3}{2}({\frac{3}{2}}^-)$, and ${\Sigma}_c\bar{D}$ states
with $\frac{3}{2}({\frac{1}{2}}^-)$, were proposed to exist within
the one-boson-exchange model~\cite{Yang:2011wz}. The two
hidden-charm pentaquark states were also found to be loosely bound
${\Sigma}_c\bar{D}^*$ and ${\Sigma}_c^*\bar{D}^*$ molecular
states, respectively, within a boson exchange interaction
model~\cite{Chen:2015loa}. Furthermore, in a meson exchange
model~\cite{He:2015cea}, $P_c(4380)$ with $J^p$=$3/2^-$ was
produced from ${\Sigma}_c^*\bar{D}$, while $P_c(4450)$ with
$J^p$=$5/2^+$ was produced from ${\Sigma}_c\bar{D}^*$. More
recently, the pentaquarks were identified with structures around
the $\Sigma_c^{(*)} \bar{D}^{(*)}$ threshold in a quark cluster
model~\cite{Takeuchi:2016ejt}.

While molecular pictures for the two pentaquark states are
quite likely, one can not rule out the possibility that these
states are compact multiquark configurations based on a strong
diquark-antidiquark pair~\cite{Maiani:2015vwa} or quark
interactions in general \cite{Takeuchi:2016nbj}. To
distinguish these two configurations, it is important
to fully explore these two possible scenarios. In this work, we will
explore the possibility that one of the pentaquark
is a compact multiquark configuration within a
constituent quark model based on the color and spin hyperfine
potential~\cite{Bhaduri:1981pn}, which is known to reproduce
the masses of the normal meson and baryon states.
In particular, in order to asses the possibility that the
$P_c(4380)$ is a compact multiquark state, we will classify the
isospin, color, and spin states for the pentaquark system
containing a heavy quark and an antiquark with $J^p$=$3/2^-$ and
isospin=1/2 from the view point of the permutation group
 which is used in characterizing a certain
symmetry so that the isospin, color, and spin states
can be represented in terms of the irreducible
Young-Yamanouchi bases. We will then systematically construct the
isospin $\otimes$ color $\otimes$ spin states satisfying the Pauli
principle from the coupling scheme appearing in the
combination of any two states. We then use the variational method
to calculate the ground state mass of the pentaquark with
$J^p$=$3/2^-$ and isospin=1/2.

This paper is organized as follows. In Sec. II, we first introduce
the Hamiltonian describing the constituent quark model, and
determine the fitting parameters of the model so as to reproduce
the mass of the baryons and mesons associated with the
thresholds.  Then, by using the variational
method, we construct the spatial wave function
suitable for a baryon and a meson. In Sec. III, we represent the
isospin, color, and spin states and then construct the
isospin $\otimes$ color $\otimes$ spin states with respect to
$I=3/2$ and $I=1/2$ in two independent basis, which can be
transformed into each other through an orthonormal matrix. We
analyze the numerical results obtained from the variational method
in Sec. IV. We finally give a summary of the paper in Sec. V.

\section{Hamiltonian}

To investigate the stability of the pentaquark in the
non-relativistic frame work, the Hamiltonian is chosen to take the
confinement and hyperfine potential for the color and spin
interaction;
\begin{eqnarray}
H=\sum_{i=1}^{5}(m_{i}+\frac{\textbf{p}^2_i}{2m_i})-\frac{3}{16}\sum_{i<j}^{4}
\lambda^c_i\lambda^c_j(V^{C}_{ij}+V^{SS}_{ij}),
\label{eq-hamiltonian}
\end{eqnarray}
where $m_i$'s are the quark masses, $\lambda^c_i/2$ the
color operator of the $i$'th quark for the color SU(3), and
$V^{C}_{ij}$ and $V^{SS}_{ij}$ the confinement and
hyperfine potential, respectively. The confinement potential is
usually composed of the linearizing term as suggested by the
lattice gauge theory, and the Coulomb-type potential as derived
from the perturbative QCD; 
\begin{align}
V^{C}_{ij}=-\frac{\kappa}{r_{ij}}+\frac{(r_{ij})^{1/2}}{a_0}-D.
\label{eq-confinement}
\end{align}
The hyperfine potential is given to take the following form,
including the spin interaction;
\begin{align}
V^{SS}_{ij}=\frac{1}{m_im_jc^4}\frac{\hbar^2c^2{\kappa}^{\prime}}{(r_{0ij})}\frac{e^{-(r_{ij})^2/(r_{0ij})^2}}{r_{ij}}
{\sigma}_i\cdot{\sigma}_j.
\label{eq-hyperfine1}
\end{align}
Here, $r_{ij}$ is the distance between interquarks,
$\mid\textbf{r}_i-\textbf{r}_j\mid$, and both
$r_{0ij}$ and ${\kappa}^{\prime}$ are
chosen to depend on the masses of interquarks, given by
\begin{align}
&r_{0ij}=1/(\alpha+\beta \frac{m_im_j}{m_i+m_j}), \nonumber \\
&{\kappa}^{\prime}={\kappa}_0(1+\gamma \frac{m_im_j}{m_i+m_j}). \label{eq-hyperfine2}
\end{align}

The hyperfine potential in Eq.~(\ref{eq-hyperfine1}), which
becomes $1/(m_im_j)$ $\delta(r)$ in the heavy quark mass limit
$m_i \to\infty$, is chosen to fit the meson and baryon mass
splitting with both light and heavy quarks. The parameters in the
Hamiltonian are fitted to the baryons and mesons masses by using
the variational method~\cite{Park:2015nha}. The fitting parameters
are given in Table~\ref{fitting-parameter}, and the calculated
masses in Table~\ref{fitting-mass}.

Since we deal with the pentaquark composed of
$q(1)q(2)q(3)c(4)\bar{c}(5)$ with $I=1/2$, where the number
indicate the position of the constituent quark, the symmetry of
the three light quarks should be taken into account to satisfy the
Pauli principle because the total wave function must be
antisymmetric among the three light quarks. As we are interested
in the ground state, a natural choice would be to
take the spatial function to be symmetric,
which requires the remaining part of the total wave function to be
antisymmetric among the three light quarks. We denote the symmetry
(antisymmetry) property by $[123]$ ($\{123\}$). In the center of
the mass frame, the pentaquark system is reduced into the
four-body problem, represented by the four Jacobian coordinates
suitable for describing the decay into a baryon and a
meson.

We take the spatial function to be a Gaussian which was
extensively used with the variational method to handle
calculations in many body problem. The four Jacobian coordinates
suitable for  describing the decay into a baryon and a
meson are given by
\begin{align}
& \pmb{x}_1^1=\frac{1}{\sqrt{2}}(\textbf{r}_1-\textbf{r}_2),\quad
\pmb{x}_2^1=\sqrt{\frac{2}{3}}(\textbf{r}_3-\frac{1}{2}\textbf{r}_1-
\frac{1}{2}\textbf{r}_2),\nonumber\\
&\pmb{x}_3^1=\frac{1}{\sqrt{2}}(\textbf{r}_4-\textbf{r}_5),\nonumber \\
&\pmb{x}_4^1=\sqrt{\frac{6}{5}}(\frac{1}{3}(\textbf{r}_1+\textbf{r}_2+\textbf{r}_3)-
\frac{1}{2}(\textbf{r}_4+\textbf{r}_5)), \label{eq-Jacobian1}
\end{align}
where the first and second terms represent a baryon configuration,
the third a meson configuration, and the last the
relative position vector between the center of mass of a baryon
and a meson. The boldface letters stand for the vectors.

\begin{widetext} 
\begin{table}[htp]
\caption{Parameters of the Hamiltonian fitted to the baryon and
meson masses occurring in the decay channels of the
$q^3c\bar{c}$.}
\begin{center}
\begin{tabular}{c|c|c|c|c|c|c|c|c}
\hline \hline
  $\gamma$  & $\kappa$  & $a_0$  & D & ${\kappa}_0$ & $\alpha$ &$\beta$ & $m_u$  & $m_c$     \\
\hline
   1.667$\rm{(GeV)^{-1}}$    &  0.107  & 1.042$\rm{(GeV)^{-2}}$    & 0.955 $\rm{GeV}$   & 0.168 $\rm{GeV}$    &  1.224 $\rm{GeV}$   & 1.467        & 0.302 $\rm{GeV}$  & 1.889 $\rm{GeV}$  \\
\hline \hline
\end{tabular}
\end{center}
\label{fitting-parameter}
\end{table}

\begin{table}[htp]
\caption{Masses of baryons and mesons obtained from the
variational method.  The third row shows the variational parameter
in $\rm{fm^{-2}}$. The fourth row shows the experimental data in
$\rm{GeV}$. }
\begin{center}
\begin{tabular}{c|c|c|c|c|c|c|c|c|c}
\hline \hline
      (I,S)   & ($\frac{1}{2}$,$\frac{1}{2}$)  P     &  ($\frac{3}{2}$,$\frac{3}{2}$) $\Delta$    &  ($0$,$\frac{1}{2}$) ${\Lambda}_c$   &  ($1$,$\frac{1}{2}$) ${\Sigma}_c$     &  ($1$,$\frac{3}{2}$) ${\Sigma}^*_c$ & ($0$,$0$)  ${\eta}_c$  & ($0$,$1$)  $J/\psi$ &  ($\frac{1}{2}$,$0$)  $D$   &  ($\frac{1}{2}$,$1$) $D^*$ \\
\hline
Mass  & 0.972  & 1.266   & 2.286       & 2.459    & 2.536   & 2.984   & 3.115  & 1.872  & 2.012   \\
Variational  &     &         &               &            &           &           &           &         &          \\
parameters& a=3.4, b=1.4  & a=2.1, b=1.2  & a=2.7, b=3.4 & a=1.9, b=3.5 & a=1.8, b=3.1 & a=15.1 & a=11 & a=4.4  & a=3.4   \\
\hline
Exp       &  0.938    &  1.232    &   2.286           & 2.453    & 2.518     & 2.983       & 3.96       & 1.869 & 2.01\\
 \hline \hline
\end{tabular}
\end{center}
\label{fitting-mass}
\end{table}
\end{widetext}

We then construct a spatial wave function given by
\begin{align}
R^{s_1}=\exp[-a_1 (\pmb{x}_1^1)^2
-a_2(\pmb{x}_2^1)^2-a_3(\pmb{x}_3^1)^2-a_4(\pmb{x}_4^1)^2],
\label{eq-spatial1}
\end{align}
where $a_1$, $a_2$, $a_3$, and $a_4$ are variational parameters.
Since the spatial wave function in Eq.~(\ref{eq-spatial1}) is
symmetric only between the particle 1 and 2, we need two
additional spatial wave functions so as to satisfy $[123]$
symmetry; one is symmetric between the particle 1 and 3, and the
other is symmetric between the particle 2 and 3. The two sets of
four Jacobian coordinates are given by
\begin{align}
&\pmb{x}_1^2=\frac{1}{\sqrt{2}}(\textbf{r}_1-\textbf{r}_3),\quad
\pmb{x}_2^2=\sqrt{\frac{2}{3}}(\textbf{r}_2-\frac{1}{2}\textbf{r}_1-
\frac{1}{2}\textbf{r}_3),\nonumber\\
&\pmb{x}_3^2=\frac{1}{\sqrt{2}}(\textbf{r}_4-\textbf{r}_5),\nonumber \\
&\pmb{x}_4^2=\sqrt{\frac{6}{5}}(\frac{1}{3}(\textbf{r}_1+\textbf{r}_2+\textbf{r}_3)-
\frac{1}{2}(\textbf{r}_4+\textbf{r}_5)), \label{eq-Jacobian2}
\end{align}

\begin{align}
&\pmb{x}_1^3=\frac{1}{\sqrt{2}}(\textbf{r}_2-\textbf{r}_3),\quad
\pmb{x}_2^3=\sqrt{\frac{2}{3}}(\textbf{r}_1-\frac{1}{2}\textbf{r}_2-
\frac{1}{2}\textbf{r}_3),\nonumber\\
&\pmb{x}_3^3=\frac{1}{\sqrt{2}}(\textbf{r}_4-\textbf{r}_5),\nonumber \\
&\pmb{x}_4^3=\sqrt{\frac{6}{5}}(\frac{1}{3}(\textbf{r}_1+\textbf{r}_2+\textbf{r}_3)-
\frac{1}{2}(\textbf{r}_4+\textbf{r}_5)), \label{eq-Jacobian3}
\end{align}
By using the two set of four Jacobian coordinates, we construct
the two spatial wave functions with either $[13]$ symmetry or
$[23]$ symmetry. Combining these spatial functions with a certain
symmetry into a linear form, we obtain the spatial function with
four variational parameters $a_1$, $a_2$, $a_3$, and $a_4$ which
is fully symmetric among the particle 1, 2, and 3 as follows;
\begin{align}
R=&\exp[-a_1(\pmb{x}_1^1)^2-a_2(\pmb{x}_2^1)^2-a_3(\pmb{x}_3^1)^2-
a_4(\pmb{x}_4^1)^2]+ \nonumber \\
&\exp[-a_1(\pmb{x}_1^2)^2-a_2(\pmb{x}_2^2)^2-a_3(\pmb{x}_3^2)^2-
a_4(\pmb{x}_4^2)^2]+\nonumber \\
&\exp[-a_1(\pmb{x}_1^3)^2-a_2(\pmb{x}_2^3)^2-a_3(\pmb{x}_3^3)^2-
a_4(\pmb{x}_4^3)^2]. \label{eq-spatial2}
\end{align}

The spatial wave function of the pentaquark in
Eq.~(\ref{eq-spatial2}) is in a state with total angular moment
$L=0$, where both the baryon and meson configurations as well as
their relative motion is in the S-wave state. The kinetic energy
part coming from Eq.~(\ref{eq-spatial2}) is given as
\begin{eqnarray}
K.E. & =& \frac{\textbf{p}^2_1+\textbf{p}^2_2}{2m_1}+
\frac{\textbf{p}^2_3}{2m_2}+ \frac{\textbf{p}^2_4}{2\mu}.
\label{eq-kinetic}
\end{eqnarray}
Here $\textbf{p}^2_1+\textbf{p}^2_2=3\hbar^2f(a_1,a_2)$,
$\textbf{p}^2_3=3\hbar^2a_3$, and  $\textbf{p}^2_4=3\hbar^2a_4$,
where $m_1,m_2$ are the light and heavy quark masses respectively,
and $\mu=5 m_1m_2/(3m_1+2m_2)$. We present $f(a_1,a_2)$ appearing
in the kinetic terms of the baryon;
\begin{align}
&f(a_1,a_2)=(a_1+a_2)\times\nonumber \\
&\{\frac{1}{(a_1a_2)^{(3/2)}}+
\frac{2048a_1a_2}{(3{a_1}^2+10a_1a_2+3{a_2}^2)^{(3/2)}}\}/ \nonumber \\
&\{\frac{2}{(a_1a_2)^{(3/2)}}+\frac{256a_1a_2}{(3{a_1}^2+10a_1a_2+3{a_2}^2)^{(3/2)}}\}.
\end{align}

Hence, for the compact mutiquark state to be stable compared to
the separated baryon and meson state, the extra attraction coming
from bringing the baryon and meson should be large enough to
overcome the extra kinetic energy given by the last term in
Eq.~(\ref{eq-kinetic}).

\section{ Isospin $\otimes$ color $\otimes$ spin state of the pentaquark}

In this section, we will construct the isospin $\otimes$ color
$\otimes$ spin state appropriate for the
$q(1)q(2)q(3)Q(4)\bar{Q}(5)$ system with $I=1/2$ and spin=3/2,
where the number in the bracket indicates the position of the
constituent quark. The component of three identical light quarks
of the pentaquark restricts the total wave function to be
antisymmetric with respect to the exchange of any pair among the
three light quarks due to Pauli principle. When the spatial
function of the pentaquark is chosen to be fully symmetric for the
three light quarks, the remaining part of the total wave function
should be fully antisymmetric. Therefore, as we are interested in
the ground state, the symmetry property of the isospin $\otimes$
color $\otimes$ spin state should be taken to be antisymmetric for
the particle 1, 2, and 3. We will use $\{123\}$ notation for the
antisymmetry property. Young tableau, which represents the
irreducible bases of the permutation group, enable us to easily
identify the multi-quark configuration with certain symmetry
property. In this paper, we will use the Young tableau and the
Young-Yamanouchi basis, which corresponds to the Young tableau in
describing the states necessary for the pentaquark. In the
following subsections, we first start by separately discussing the
isospin, color and spin states consisting of five quarks, and then
discuss the total wave function.

\subsection{Isospin states}

In the SU(2) flavor symmetry, it is easy to find that the possible
isospin ($I$) states for the three light quarks are 1/2 and 3/2.
The Young-Yamanouchi basis corresponding to the $I=1/2$ state is
as follows:
\begin{align}
&\begin{tabular}{c}
$\vert I^{1/2}_1 \rangle$=
\end{tabular}
\begin{tabular}{|c|c|}
\hline
1 & 2   \\
\cline{1-2}
\multicolumn{1}{|c|}{3}  \\
\cline{1-1}
\end{tabular}
=\frac{1}{\sqrt{6}}(2uud-udu-duu),
\nonumber
\\
&\begin{tabular}{c}
$\vert I^{1/2}_2 \rangle$=
\end{tabular}
\begin{tabular}{|c|c|}
\hline
1 & 3  \\
\cline{1-2}
\multicolumn{1}{|c|}{2}  \\
\cline{1-1}
\end{tabular}
=\frac{1}{\sqrt{2}}(udu-duu).
\label{eq-flavor2}
\end{align}
\subsection{Color states}
For the possible color states, we only consider the color singlets
which are assumed to be observables in hadron state. There are
several ways of obtaining the color singlets for the pentaquark,
coming from the direct product, given by
\begin{align}
&[3]_C\otimes[3]_C\otimes[3]_C\otimes[3]_C\otimes\bar{[3]}_C. \nonumber
\end{align}
We introduce the two methods which are equivalent to each other,
but different in the way of combining the irreducible
representation of SU(3). First, since the antiquark corresponds to
the antitriplet, we can construct the triplet in the direct
product, $[3]_C\otimes[3]_C\otimes[3]_C\otimes[3]_C$, which
corresponds to Young tableau [2,1,1];
\begin{align}
&\begin{tabular}{|c|c|}
\hline
           1 &  2    \\
\cline{1-2}
\multicolumn{1}{|c|}{3} \\
\cline{1-1}
\multicolumn{1}{|c|}{4}  \\
\cline{1-1}
\end{tabular}
=\{(12)_6(34)_{\bar{3}}\}_3,
\quad
\begin{tabular}{|c|c|}
\hline
           1 &  3    \\
\cline{1-2}
\multicolumn{1}{|c|}{2} \\
\cline{1-1}
\multicolumn{1}{|c|}{4}  \\
\cline{1-1}
\end{tabular}
=\{(12)_634\}_3,
\nonumber \\
&\begin{tabular}{|c|c|}
\hline
           1 &  4    \\
\cline{1-2}
\multicolumn{1}{|c|}{2} \\
\cline{1-1}
\multicolumn{1}{|c|}{3}  \\
\cline{1-1}
\end{tabular}
=\{(123)_14\}_3.
\label{eq-color1}
\end{align}
Here, the subscript indicates the irreducible representation of
SU(3). Then, we can obtain the three color singlets,
combining the triplet in Eq.~(\ref{eq-color1}) with the
antitriplet of antiquark. We denote the color singlets by,
\begin{align}
&\vert C_1 \rangle=
\begin{tabular}{|c|c|}
\hline
           1 &  2    \\
\cline{1-2}
\multicolumn{1}{|c|}{3} \\
\cline{1-1}
\multicolumn{1}{|c|}{4}  \\
\cline{1-1}
\end{tabular}_3
\otimes(5)_{\bar{3}},
\quad
\vert C_2 \rangle=
\begin{tabular}{|c|c|}
\hline
           1 &  3    \\
\cline{1-2}
\multicolumn{1}{|c|}{2} \\
\cline{1-1}
\multicolumn{1}{|c|}{4}  \\
\cline{1-1}
\end{tabular}_3
\otimes(5)_{\bar{3}},
\nonumber \\
&\vert C_3 \rangle=
\begin{tabular}{|c|c|}
\hline
           1 &  4    \\
\cline{1-2}
\multicolumn{1}{|c|}{2} \\
\cline{1-1}
\multicolumn{1}{|c|}{3}  \\
\cline{1-1}
\end{tabular}_3
\otimes(5)_{\bar{3}}.
\label{eq-color2}
\end{align}
Secondly, we can decompose the direct product,
$[3]_C\otimes[3]_C\otimes[3]_C$ and
$\otimes[3]_C\otimes\bar{[3]}_C$ into the direct sum of the
irreducible representations, respectively, as follows;
\begin{align}
&[3]_C\otimes[3]_C\otimes[3]_C=
\begin{tabular}{|c|c|}
\hline
1 & 2   \\
\cline{1-2}
\multicolumn{1}{|c|}{3}  \\
\cline{1-1}
\end{tabular}_8
 \oplus
\begin{tabular}{|c|c|}
\hline
           1 &  3    \\
\cline{1-2}
\multicolumn{1}{|c|}{2} \\
\cline{1-1}
\end{tabular}_8
 \oplus
\begin{tabular}{|c|}
\hline
           1    \\
\hline
2             \\
\hline
3             \\
\hline
\end{tabular}_{ 1},
\label{eq-color4} \\
&[3]_C\otimes\bar{[3]}_C=[8]_C \oplus[1]_C.
\label{eq-color5}
\end{align}
Then, the coupling of either the octet with the octet or the
singlet with the singlet in Eq.~(\ref{eq-color4}) and
Eq.~(\ref{eq-color5}) gives the three color singlets of the
pentaquark, denoted by,
\begin{align}
&\vert C_1 \rangle=
\begin{tabular}{|c|c|}
\hline
           1 &  2    \\
\cline{1-2}
\multicolumn{1}{|c|}{3} \\
\cline{1-1}
\end{tabular}_8
\otimes(45)_{8},
\quad
\vert C_2 \rangle=
\begin{tabular}{|c|c|}
\hline
           1 &  3    \\
\cline{1-2}
\multicolumn{1}{|c|}{2} \\
\cline{1-1}
\end{tabular}_8
\otimes(45)_{8},
\nonumber \\
&\vert C_3 \rangle=
\begin{tabular}{|c|}
\hline
           1 \\
\hline
           2 \\
\hline
           3 \\
 \hline
\end{tabular}_1
\otimes(45)_{1}.
\label{eq-color6}
\end{align}
It should be noted that the color singlets represented in terms of
different Young tableau in Eq.~(\ref{eq-color2}) and
Eq.~(\ref{eq-color6}) are the same in a tensor form. We define the
color singlets derived from the above methods, as follows;
\begin{align}
&\vert C_1 \rangle=[\{(12)_6(34)_{\bar{3}}\}_{3}5_{\bar{3}}]_{1}=[\{(12)_63\}_8(45)_8]_{1}, \nonumber \\
&\vert C_2 \rangle=[\{(12)_{\bar{3}}34\}_{3}5_{\bar{3}}]_{1}=[\{(12)_{\bar{3}}3\}_8(45)_8]_{1}, \nonumber \\
&\vert C_3 \rangle=[\{(123)_14\}_35_{\bar{3}}]_{1}=[\{(123)_1(45)_{1}]_{1}.
\label{eq-color7}
\end{align}

\subsection{Spin states}

For the spin=3/2 pentaquark case, the spin states are represented
in terms of Young tableau [4,1] with four dimension, as follows:
\begin{align}
&\vert S^{3/2}_1 \rangle=
\begin{tabular}{|c|c|c|c|}
\hline
1& 2 &  3& 4  \\
\cline{1-4}
\multicolumn{1}{|c|}{5} \\
\cline{1-1}
\end{tabular},
\vert S^{3/2}_2 \rangle=
\begin{tabular}{|c|c|c|c|}
\hline
1& 2 &  3& 5  \\
\cline{1-4}
\multicolumn{1}{|c|}{4} \\
\cline{1-1}
\end{tabular},
\vert S^{3/2}_3 \rangle=
\begin{tabular}{|c|c|c|c|}
\hline
1& 2 &  4& 5  \\
\cline{1-4}
\multicolumn{1}{|c|}{3} \\
\cline{1-1}
\end{tabular},
\nonumber \\
&\vert S^{3/2}_4 \rangle=
\begin{tabular}{|c|c|c|c|}
\hline
1& 3 &  4& 5  \\
\cline{1-4}
\multicolumn{1}{|c|}{2} \\
\cline{1-1}
\end{tabular}.
\label{eq-spin1}
\end{align}
When we investigate the stability of the pentaquark against the
strong decay into a baryon and a meson, it is very convenient to
use the spin states related with the decay mode. We denote the
four spin states by,
\begin{align}
&\vert{\phi}_1 \rangle=[\{(12)_13_{1/2}\}_{3/2}(45)_0]_{3/2}, \nonumber \\
&\vert{\phi}_2 \rangle=[\{(12)_13_{1/2}\}_{3/2}(45)_1]_{3/2}, \nonumber \\
&\vert{\phi}_3 \rangle=[\{(12)_13_{1/2}\}_{1/2}(45)_1]_{3/2}, \nonumber \\
&\vert{\phi}_4 \rangle=[\{(12)_03_{1/2}\}_{1/2}(45)_1]_{3/2},
\label{eq-spin2}
\end{align}
where the subscript indicates the spin state. Due to the
orthonormality of the two sets of spin sates, Eq.~(\ref{eq-spin1})
and  Eq.~(\ref{eq-spin2}) are related by the following orthogonal
transformation:
\begin{align}
\left(\begin{array}{cccc} \sqrt{\frac{5}{8}}  &   \sqrt{\frac{3}{8}} & 0 & 0 \\
                         -\sqrt{\frac{3}{8}}  &   \sqrt{\frac{5}{8}} & 0 & 0         \\
                                          0 &  0 & 1 & 0         \\
                            0 &  0 & 0 & 1    \end{array} \right).
\label{eq-spin3}
\end{align}

\subsection{Isospin $\otimes$ color $\otimes$ spin state for $I=1/2$}

Since the isospin, color and spin states represented in terms of
the Young tableau have a certain symmetry property, we can
construct the isospin $\otimes$ color $\otimes$ spin state of the
pentaquark which is fully antisymmetric under the exchange of any
pair among the particle 1, 2 and 3. For this purpose,
depending on how the coupling scheme is implemented, we
consider two methods. In the first method, we start from the
 notation of the color singlets in
Eq.~(\ref{eq-color2}), and combine the color singlets with spin
states by the out product of the permutation group, $S_4$,
resulting in the color $\otimes$ spin states for the particle 1,
2, 3, and 4. Then, we can easily obtain the isospin $\otimes$
color $\otimes$ spin state with $\{123\}$ symmetry by coupling of
the isospin state with the color $\otimes$ spin states. In the
second method, we start the notation of the color singlets in
Eq.~(\ref{eq-color6}), and use the $S_3$ permutation group
 applied on the coupling scheme.

According to the permutation group theory~\cite{Chen:2002gd}, the
irreducible basis of $S_5$ becomes the irreducible basis of $S_4$
as well, irrespective of the particle 5. When we consider the
symmetry property for the particle 1, 2, 3, and 4 in coupling
scheme, we can identify the spin states in Eq.~(\ref{eq-spin1})
with the Young-Yamanouchi bases for Young tableau [4] and Young
tableau [3,1] without the particle 5;
\begin{align}
&\vert S^{3/2}_1 \rangle=
\begin{tabular}{|c|c|c|c|}
\hline
1& 2 &  3& 4  \\
\hline
\end{tabular},
\vert S^{3/2}_2 \rangle=
\begin{tabular}{|c|c|c|}
\hline
1& 2 &  3 \\
\cline{1-3}
\multicolumn{1}{|c|}{4} \\
\cline{1-1}
\end{tabular},
\vert S^{3/2}_3 \rangle=
\begin{tabular}{|c|c|c|}
\hline
1& 2 &  4  \\
\cline{1-3}
\multicolumn{1}{|c|}{3} \\
\cline{1-1}
\end{tabular},
\nonumber \\
&\vert S^{3/2}_4 \rangle=
\begin{tabular}{|c|c|c|}
\hline
1& 3 &  4  \\
\cline{1-3}
\multicolumn{1}{|c|}{2} \\
\cline{1-1}
\end{tabular}.
\label{eq-spin4}
\end{align}
It is necessary to show the outer product between Young
tableau [2,1,1] of the color singlets in Eq.~(\ref{eq-color2}) and
Young tableau [3,1] of the spin states in Eq.~(\ref{eq-spin4}) so
that we obtain the color $\otimes$ spin states;
\begin{align}
\begin{tabular}{|c|c|}
\hline
   $\quad$         &  $\quad$   \\
\cline{1-2}
\multicolumn{1}{|c|}{$\quad$} \\
\cline{1-1}
\multicolumn{1}{|c|}{$\quad$}  \\
\cline{1-1}
\end{tabular}_C
 \otimes
 \begin{tabular}{|c|c|c|}
\hline
 $\quad$   &  $\quad$    &   $\quad$    \\
\cline{1-3}
\multicolumn{1}{|c|}{ $\quad$  } \\
\cline{1-1}
\end{tabular}_S
=
&\begin{tabular}{|c|}
\hline
 $\quad$      \\
\hline
 $\quad$      \\
 \hline
 $\quad$      \\
 \hline
 $\quad$      \\
\hline
\end{tabular}_{CS_1}
\oplus
\begin{tabular}{|c|c|}
\hline
   $\quad$         &  $\quad$   \\
\cline{1-2}
\multicolumn{1}{|c|}{$\quad$} \\
\cline{1-1}
\multicolumn{1}{|c|}{$\quad$}  \\
\cline{1-1}
\end{tabular}_{CS_2}
\oplus
 \begin{tabular}{|c|c|}
\hline
 $\quad$   &  $\quad$       \\
\hline
 $\quad$   &  $\quad$       \\
\hline
\end{tabular}_{CS_3}
\oplus
\nonumber \\
&\begin{tabular}{|c|c|c|}
\hline
 $\quad$   &  $\quad$    &   $\quad$    \\
\cline{1-3}
\multicolumn{1}{|c|}{ $\quad$  } \\
\cline{1-1}
\end{tabular}_{CS_4}
.
\label{eq-cs1}
\end{align}
In addition to this, we should consider the outer product between
Young tableau [2,1,1] of the color singlets in
Eq.~(\ref{eq-color2}) and Young tableau [4] of the spin states in
Eq.~(\ref{eq-spin4});
\begin{align}
\begin{tabular}{|c|c|}
\hline
   $\quad$         &  $\quad$   \\
\cline{1-2}
\multicolumn{1}{|c|}{$\quad$} \\
\cline{1-1}
\multicolumn{1}{|c|}{$\quad$}  \\
\cline{1-1}
\end{tabular}_C
 \otimes
 \begin{tabular}{|c|c|c|c|}
\hline
 $\quad$   &  $\quad$    &   $\quad$ &  $\quad$ \\
\hline
\end{tabular}_S
=
\begin{tabular}{|c|c|}
\hline
   $\quad$         &  $\quad$   \\
\cline{1-2}
\multicolumn{1}{|c|}{$\quad$} \\
\cline{1-1}
\multicolumn{1}{|c|}{$\quad$}  \\
\cline{1-1}
\end{tabular}_{CS_5}
.
\label{eq-cs2}
\end{align}

The coupling scheme designed to construct the isospin $\otimes$
color $\otimes$ spin states with the $\{123\}$ symmetry is
completed by using the Clebsch-Gordan (CG) coefficient of the
permutation group, $S_n$, which is factorized into the
Clebsch-Gordan (CG) coefficient of $S_{n-1}$ and K
matrix~\cite{Stancu:1999qr}, given by,
\begin{align}
&S([f^{\prime}]p^{\prime}q^{\prime}y^{\prime}[f^{\prime\prime}]p^{\prime\prime}q^{\prime\prime}y^{\prime\prime}\vert[f]pqy)=\nonumber\\
&K([f^{\prime}]p^{\prime}[f^{\prime\prime}]p^{\prime\prime}\vert[f]p)
S([f^{\prime}_{p^{\prime}}]q^{\prime}y^{\prime}[f^{\prime\prime}_{p^{\prime\prime}}]q^{\prime\prime}y^{\prime\prime}\vert[f_p]qy),
\label{K-matrix}
\end{align}
where $S$ in the left-hand (right-hand) side is a CG
coefficient of $S_n$ ($S_{n-1}$). In this work, we take a
similar process which was described in Refs~\cite{Park:2015nha,Park:2016cmg}.

Below, we show the Young-Yamanouchi bases corresponding to Young
tableau [2,1,1] which is obtained from the color $\otimes$ spin
coupling in Eq.~(\ref{eq-cs1});
\begin{align}
\begin{tabular}{|c|c|}
\hline
           1 &  2    \\
\cline{1-2}
\multicolumn{1}{|c|}{3} \\
\cline{1-1}
\multicolumn{1}{|c|}{4}  \\
\cline{1-1}
\end{tabular}_{CS_2}
=&-\frac{1}{\sqrt{6}}\vert C_1 \rangle \otimes \vert S^{3/2}_2 \rangle
-\frac{1}{\sqrt{3}}\vert C_1 \rangle \otimes \vert S^{3/2}_3 \rangle
\nonumber \\
&+\frac{1}{\sqrt{3}}\vert C_2 \rangle \otimes \vert S^{3/2}_4 \rangle
-\frac{1}{\sqrt{6}}\vert C_3 \rangle \otimes \vert S^{3/2}_4 \rangle.
\end{align}
\begin{align}
\begin{tabular}{|c|c|}
\hline
           1 &  3    \\
\cline{1-2}
\multicolumn{1}{|c|}{2} \\
\cline{1-1}
\multicolumn{1}{|c|}{4}  \\
\cline{1-1}
\end{tabular}_{CS_2}
=&\frac{1}{\sqrt{3}}\vert C_1 \rangle \otimes \vert S^{3/2}_4 \rangle
-\frac{1}{\sqrt{6}}\vert C_2 \rangle \otimes \vert S^{3/2}_2 \rangle
\nonumber \\
&+\frac{1}{\sqrt{3}}\vert C_2 \rangle \otimes \vert S^{3/2}_3 \rangle
+\frac{1}{\sqrt{6}}\vert C_3 \rangle \otimes \vert S^{3/2}_3 \rangle.
\end{align}
For the case of Young tableau [2,2], which is obtained from the
color $\otimes$ spin coupling in Eq.~(\ref{eq-cs1}), the
Young-Yamanouchi bases are as follows;
\begin{align}
\begin{tabular}{|c|c|}
\hline
           1 &  2    \\
\hline
           3 &  4    \\
\hline
\end{tabular}_{CS_3}
=&-\frac{1}{\sqrt{3}}\vert C_1 \rangle \otimes \vert S^{3/2}_2 \rangle
+\frac{1}{\sqrt{6}}\vert C_1 \rangle \otimes \vert S^{3/2}_3 \rangle
\nonumber \\
&-\frac{1}{\sqrt{6}}\vert C_2 \rangle \otimes \vert S^{3/2}_4 \rangle
-\frac{1}{\sqrt{3}}\vert C_3 \rangle \otimes \vert S^{3/2}_4 \rangle.
\end{align}
\begin{align}
\begin{tabular}{|c|c|}
\hline
           1 &  3   \\
\hline
           2 &  4    \\
\hline
\end{tabular}_{CS_3}
=&-\frac{1}{\sqrt{6}}\vert C_1 \rangle \otimes \vert S^{3/2}_4 \rangle
-\frac{1}{\sqrt{3}}\vert C_2 \rangle \otimes \vert S^{3/2}_2 \rangle
\nonumber \\
&-\frac{1}{\sqrt{6}}\vert C_2 \rangle \otimes \vert S^{3/2}_3 \rangle
+\frac{1}{\sqrt{3}}\vert C_3 \rangle \otimes \vert S^{3/2}_3 \rangle.
\end{align}
For the case of Young tableau [3,1], which is obtained from the
color $\otimes$ spin coupling in Eq.~(\ref{eq-cs1}),
Young-Yamanouchi bases are as follows;
\begin{align}
\begin{tabular}{|c|c|c|}
\hline
           1 &  2 & 4   \\
\cline{1-3}
\multicolumn{1}{|c|}{3} \\
\cline{1-1}
\end{tabular}_{CS_4}
=&-\frac{1}{\sqrt{2}}\vert C_1 \rangle \otimes \vert S^{3/2}_2 \rangle
+\frac{1}{\sqrt{2}}\vert C_3 \rangle \otimes \vert S^{3/2}_4 \rangle.
\end{align}
\begin{align}
\begin{tabular}{|c|c|c|}
\hline
           1 &  3 & 4   \\
\cline{1-3}
\multicolumn{1}{|c|}{2} \\
\cline{1-1}
\end{tabular}_{CS_4}
=&-\frac{1}{\sqrt{2}}\vert C_2 \rangle \otimes \vert S^{3/2}_2 \rangle
-\frac{1}{\sqrt{2}}\vert C_3 \rangle \otimes \vert S^{3/2}_3 \rangle.
\end{align}
For the case of Young tableau [2,1,1], which is obtained from the
color $\otimes$ spin coupling in Eq.~(\ref{eq-cs2}), the
Young-Yamanouchi bases are as follows;
\begin{align}
\begin{tabular}{|c|c|}
\hline
           1 &  2    \\
\cline{1-2}
\multicolumn{1}{|c|}{3} \\
\cline{1-1}
\multicolumn{1}{|c|}{4}  \\
\cline{1-1}
\end{tabular}_{CS_5}
=\vert C_1 \rangle \otimes \vert S^{3/2}_1 \rangle.
\end{align}
\begin{align}
\begin{tabular}{|c|c|}
\hline
           1 &  3    \\
\cline{1-2}
\multicolumn{1}{|c|}{2} \\
\cline{1-1}
\multicolumn{1}{|c|}{4}  \\
\cline{1-1}
\end{tabular}_{CS_5}
=\vert C_2 \rangle \otimes \vert S^{3/2}_1 \rangle.
\end{align}

To find the isospin $\otimes$ color $\otimes$ spin state with
$\{123\}$ symmetry, we finally combine the isospin states in
Eq.~(\ref{eq-flavor2}) with color $\otimes$ spin states for Young
tableau [2,1,1] in Eq.~(\ref{eq-cs2}) as well as Young tableau
[2,1,1], [2,2], and [3,1] in Eq.~(\ref{eq-cs1}). Therefore, we
have four isospin $\otimes$ color $\otimes$ spin states with
$\{123\}$ symmetry for $I=1/2$;
\begin{align}
&\vert [I^{\frac{1}{2}}CS]_1 \rangle=\frac{1}{\sqrt{2}}\Big(
\begin{tabular}{|c|c|}
\hline
1 & 2   \\
\cline{1-2}
\multicolumn{1}{|c|}{3}  \\
\cline{1-1}
\end{tabular}_I
\otimes
\begin{tabular}{|c|c|}
\hline
           1 &  3    \\
\cline{1-2}
\multicolumn{1}{|c|}{2} \\
\cline{1-1}
\multicolumn{1}{|c|}{4}  \\
\cline{1-1}
\end{tabular}_{CS_2}
-
\begin{tabular}{|c|c|}
\hline
1 & 3   \\
\cline{1-2}
\multicolumn{1}{|c|}{2}  \\
\cline{1-1}
\end{tabular}_I
\otimes
\begin{tabular}{|c|c|}
\hline
           1 &  2    \\
\cline{1-2}
\multicolumn{1}{|c|}{3} \\
\cline{1-1}
\multicolumn{1}{|c|}{4}  \\
\cline{1-1}
\end{tabular}_{CS_2}\Big)
\nonumber \\
&\vert [I^{\frac{1}{2}}CS]_2 \rangle=\frac{1}{\sqrt{2}}\Big(
\begin{tabular}{|c|c|}
\hline
1 & 2   \\
\cline{1-2}
\multicolumn{1}{|c|}{3}  \\
\cline{1-1}
\end{tabular}_I
\otimes
\begin{tabular}{|c|c|}
\hline
           1 &  3    \\
\hline
           2 &  4    \\
\hline
\end{tabular}_{CS_3}
-
\begin{tabular}{|c|c|}
\hline
1 & 3   \\
\cline{1-2}
\multicolumn{1}{|c|}{2}  \\
\cline{1-1}
\end{tabular}_I
\otimes
\begin{tabular}{|c|c|}
\hline
           1 &  2    \\
\hline
           3 &  4    \\
\hline
\end{tabular}_{CS_3}\Big)
\nonumber \\
&\vert [I^{\frac{1}{2}}CS]_3 \rangle=\frac{1}{\sqrt{2}}\Big(
\begin{tabular}{|c|c|}
\hline
1 & 2   \\
\cline{1-2}
\multicolumn{1}{|c|}{3}  \\
\cline{1-1}
\end{tabular}_I
\otimes
\begin{tabular}{|c|c|c|}
\hline
           1 &  3 & 4    \\
\cline{1-3}
\multicolumn{1}{|c|}{2} \\
\cline{1-1}
\end{tabular}_{CS_4}
-
\begin{tabular}{|c|c|}
\hline
1 & 3   \\
\cline{1-2}
\multicolumn{1}{|c|}{2}  \\
\cline{1-1}
\end{tabular}_I
\otimes
\begin{tabular}{|c|c|c|}
\hline
           1 &  2 & 4   \\
\cline{1-3}
\multicolumn{1}{|c|}{3} \\
\cline{1-1}
\end{tabular}_{CS_4}\Big)
\nonumber \\
&\vert [I^{\frac{1}{2}}CS]_4 \rangle=\frac{1}{\sqrt{2}}\Big(
\begin{tabular}{|c|c|}
\hline
1 & 2   \\
\cline{1-2}
\multicolumn{1}{|c|}{3}  \\
\cline{1-1}
\end{tabular}_I
\otimes
\begin{tabular}{|c|c|}
\hline
           1 &  3    \\
\cline{1-2}
\multicolumn{1}{|c|}{2} \\
\cline{1-1}
\multicolumn{1}{|c|}{4}  \\
\cline{1-1}
\end{tabular}_{CS_5}
-
\begin{tabular}{|c|c|}
\hline
1 & 3   \\
\cline{1-2}
\multicolumn{1}{|c|}{2}  \\
\cline{1-1}
\end{tabular}_I
\otimes
\begin{tabular}{|c|c|}
\hline
           1 &  2   \\
\cline{1-2}
\multicolumn{1}{|c|}{3} \\
\cline{1-1}
\multicolumn{1}{|c|}{4}  \\
\cline{1-1}
\end{tabular}_{CS_5}\Big)
.
\label{eq-ICS1-I-1/2}
\end{align}

From the notation of the color singlets in Eq.~(\ref{eq-color6})
which represents the symmetry of the permutation group, $S_3$, we
easily see that the $\vert C_3 \rangle$ state
has the symmetry property with $\{123\}$. For
that reason, the isospin $\otimes$ spin state in combining with
the $\vert C_3 \rangle$ state should be fully symmetric in the
exchange of any pair among the particle 1, 2, and 3, and the
coupling of $\vert C_3 \rangle$ state with the isospin $\otimes$
spin states gives the isospin $\otimes$ color $\otimes$ spin state
with $\{123\}$ symmetry. We denote the isospin $\otimes$ spin
states satisfying fully symmetry, by
\begin{align}
\begin{tabular}{|c|c|c|}
\hline
1 & 2 &3  \\
\hline
\end{tabular}_{IS}
=
\frac{1}{\sqrt{2}}\Big(
\begin{tabular}{|c|c|}
\hline
1 & 2   \\
\cline{1-2}
\multicolumn{1}{|c|}{3}  \\
\cline{1-1}
\end{tabular}_I
\otimes
\begin{tabular}{|c|c|c|c|}
\hline
1& 2 &  4& 5  \\
\cline{1-4}
\multicolumn{1}{|c|}{3} \\
\cline{1-1}
\end{tabular}_S
+
\begin{tabular}{|c|c|}
\hline
1 & 3  \\
\cline{1-2}
\multicolumn{1}{|c|}{2}  \\
\cline{1-1}
\end{tabular}_I
\otimes
\begin{tabular}{|c|c|c|c|}
\hline
1& 3 &  4& 5  \\
\cline{1-4}
\multicolumn{1}{|c|}{2} \\
\cline{1-1}
\end{tabular}_S\Big).
\label{eq-IS}
\end{align}
On the contrary, since both $\vert S^{3/2}_1 \rangle$ and $\vert
S^{3/2}_2 \rangle$ sates in  Eq.~(\ref{eq-spin1}) are fully
symmetric in the exchange of any pair among the particle 1, 2, and
3, the isospin $\otimes$ color state in combining with either
$\vert S^{3/2}_1 \rangle$ or $\vert S^{3/2}_2 \rangle$ state
should have the opposite symmetry due to the same reason. We
denote the isospin $\otimes$ color state satisfying fully
antisymmetry, by
\begin{align}
\begin{tabular}{|c|}
\hline
1  \\
\hline
2  \\
\hline
3  \\
\hline
\end{tabular}_{IC}
=
\frac{1}{\sqrt{2}}\Big(
\begin{tabular}{|c|c|}
\hline
1 & 2   \\
\cline{1-2}
\multicolumn{1}{|c|}{3}  \\
\cline{1-1}
\end{tabular}_I
\otimes
\begin{tabular}{|c|c|}
\hline
           1 &  3    \\
\cline{1-2}
\multicolumn{1}{|c|}{2} \\
\cline{1-1}
\end{tabular}_8
\otimes(45)_{8}
-
\begin{tabular}{|c|c|}
\hline
1 & 3   \\
\cline{1-2}
\multicolumn{1}{|c|}{2}  \\
\cline{1-1}
\end{tabular}_I
\otimes
\begin{tabular}{|c|c|}
\hline
           1 &  2    \\
\cline{1-2}
\multicolumn{1}{|c|}{3} \\
\cline{1-1}
\end{tabular}_8
\otimes(45)_{8}
\Big).
\label{eq-IC}
\end{align}
Lastly, we can consider the color $\otimes$ spin states
corresponding to Young tableau which are conjugate to that of the
isospin states, for the reason why any fully antisymmetric state
can be obtained by the coupling of any Young tableau with the
conjugate. We denote the color $\otimes$ spin states corresponding
to Young tableau [2,1] for the particle 1, 2, and 3, by
\begin{align}
\begin{tabular}{|c|c|}
\hline
1 & 2   \\
\cline{1-2}
\multicolumn{1}{|c|}{3}  \\
\cline{1-1}
\end{tabular}_{CS}
=
\frac{1}{\sqrt{2}}\Big(
&\begin{tabular}{|c|c|}
\hline
           1 &  2    \\
\cline{1-2}
\multicolumn{1}{|c|}{3} \\
\cline{1-1}
\end{tabular}_8
\otimes(45)_{8}
\otimes
\begin{tabular}{|c|c|c|c|}
\hline
1& 2 &  4& 5  \\
\cline{1-4}
\multicolumn{1}{|c|}{3} \\
\cline{1-1}
\end{tabular}_S
-
\nonumber \\
&\begin{tabular}{|c|c|}
\hline
           1 &  3    \\
\cline{1-2}
\multicolumn{1}{|c|}{2} \\
\cline{1-1}
\end{tabular}_8
\otimes(45)_{8}
\otimes
\begin{tabular}{|c|c|c|c|}
\hline
1& 3 &  4& 5  \\
\cline{1-4}
\multicolumn{1}{|c|}{2} \\
\cline{1-1}
\end{tabular}_S
\Big),
\nonumber \\
\begin{tabular}{|c|c|}
\hline
1 & 3   \\
\cline{1-2}
\multicolumn{1}{|c|}{3}  \\
\cline{1-1}
\end{tabular}_{CS}
=
-\frac{1}{\sqrt{2}}\Big(
&\begin{tabular}{|c|c|}
\hline
           1 &  2    \\
\cline{1-2}
\multicolumn{1}{|c|}{3} \\
\cline{1-1}
\end{tabular}_8
\otimes(45)_{8}
\otimes
\begin{tabular}{|c|c|c|c|}
\hline
1& 3 &  4& 5  \\
\cline{1-4}
\multicolumn{1}{|c|}{2} \\
\cline{1-1}
\end{tabular}_S
+
\nonumber \\
&\begin{tabular}{|c|c|}
\hline
           1 &  3    \\
\cline{1-2}
\multicolumn{1}{|c|}{2} \\
\cline{1-1}
\end{tabular}_8
\otimes(45)_{8}
\otimes
\begin{tabular}{|c|c|c|c|}
\hline
1& 2 &  4& 5  \\
\cline{1-4}
\multicolumn{1}{|c|}{3} \\
\cline{1-1}
\end{tabular}_S
\Big).
\label{eq-CS}
\end{align}
We denote another set of the isospin $\otimes$ color $\otimes$
spin states satisfying fully symmetry, by
\begin{align}
&\vert \psi_1 \rangle=
\begin{tabular}{|c|}
\hline
           1 \\
\hline
           2 \\
\hline
           3 \\
 \hline
\end{tabular}_1
\otimes(45)_{1}
\otimes
\begin{tabular}{|c|c|c|}
\hline
    1 & 2 & 3 \\
\hline
\end{tabular}_{IS}
,
\vert \psi_2 \rangle=
\begin{tabular}{|c|}
\hline
           1    \\
\hline
           2    \\
\hline
           3    \\
\hline
\end{tabular}_{IC}
 \otimes
\begin{tabular}{|c|c|c|c|}
\hline
1& 2 &  3& 5  \\
\cline{1-4}
\multicolumn{1}{|c|}{4} \\
\cline{1-1}
\end{tabular}_S
 , \nonumber \\
&\vert \psi_3 \rangle=
\frac{1}{\sqrt{2}}\Big(
\begin{tabular}{|c|c|}
\hline
           1 &  2    \\
\cline{1-2}
\multicolumn{1}{|c|}{3} \\
\cline{1-1}
\end{tabular}_I
\otimes
\begin{tabular}{|c|c|}
\hline
           1 &  3    \\
\cline{1-2}
\multicolumn{1}{|c|}{2} \\
\cline{1-1}
\end{tabular}_{CS}
-
\begin{tabular}{|c|c|}
\hline
           1 &  3    \\
\cline{1-2}
\multicolumn{1}{|c|}{2} \\
\cline{1-1}
\end{tabular}_I
\otimes
\begin{tabular}{|c|c|}
\hline
           1 &  2    \\
\cline{1-2}
\multicolumn{1}{|c|}{3} \\
\cline{1-1}
\end{tabular}_{CS}\Big)
,
\nonumber  \\
&\vert \psi_4 \rangle=
\begin{tabular}{|c|}
\hline
           1    \\
\hline
           2    \\
\hline
           3    \\
\hline
\end{tabular}_{IC}
 \otimes
\begin{tabular}{|c|c|c|c|}
\hline
1& 2 &  3& 4  \\
\cline{1-4}
\multicolumn{1}{|c|}{5} \\
\cline{1-1}
\end{tabular}_S
.
\label{eq-ICS2-I-1/2}
\end{align}
We note that both the states in Eq.~(\ref{eq-ICS1-I-1/2}) and the
states in Eq.~(\ref{eq-ICS2-I-1/2}) are orthonormal to each other
in four dimension vector space, respectively.

It is worthwhile to mention  that from a hadron state point of
view $\vert \psi_1 \rangle$ accounts for the $(p)_1 \otimes
(J/\psi)_1$ state, where the subscript indicates the color state,
in a fact that the color part consists of the color singlet of a
baryon multiplied by the color singlet of a meson, and the spin
part contains a baryon with spin=1/2 multiplied by a meson with
spin=1 in Eq.~(\ref{eq-spin3}). On the other hand, $\vert \psi_2
\rangle$ represents the $(p)_8 \otimes (J/\psi)_8$ state as an
unphysical state, since the color singlet represents the hidden
color, coming from the color octet of a baryon multiplied by the
color octet of a meson. The rest corresponds to a unphysical
state, resulting from the property of the pentaquark and the Pauli
principle.

In a vector space with four dimension where the isospin $\otimes$
color $\otimes$ spin states have the symmetry property with
$\{123\}$, there exists orthogonal matrix which transforms the set
of Eq.~(\ref{eq-ICS2-I-1/2}) into the set of
Eq.~(\ref{eq-ICS1-I-1/2}),  given by,
\begin{align}
\left(\begin{array}{cccc} \frac{1}{\sqrt{6}}  &   \frac{1}{\sqrt{3}} &  -\frac{1}{\sqrt{2}} & 0 \\
                        -\frac{1}{\sqrt{6}}  &   -\frac{1}{\sqrt{3}} &  -\frac{1}{\sqrt{2}} & 0        \\
                             -\frac{\sqrt{2}}{\sqrt{3}}  &   -\frac{1}{\sqrt{3}} &  0            & 0    \\
                            0         &  0        & 0                       & 1                \end{array} \right).
\label{eq-transform-ICS1}
\end{align}

\section{Numerical Results}

In this section, we analyze the numerical results performed using
the variational method for the Hamiltonian given in
Eq.~(\ref{eq-hamiltonian}). For that purpose, we adopt the trial
wave function which consists of the spatial function in
Eq.~(\ref{eq-spatial2}) and the isospin $\otimes$ color $\otimes$
spin states obtained from Sec. III. The trial wave function can
thus be expanded as follows:
\begin{align}
\vert {\Psi}_{\alpha}\rangle= \sum_{i}C^{\alpha}_i \vert R \rangle \vert [ICS]_i \rangle.
\label{eq-trial}
\end{align}

Before discussing the numerical analysis, it is useful to examine
the expectation value of the color spin part of the hyperfine
potential, with the spatial dependence factored out, in the matrix
form generated by the four independent isospin $\otimes$ color
$\otimes$ spin states. This hyperfine matrix is essential in
identifying possible attraction in the four configurations. A
stable or resonant pentaquark state can only exist if the
hyperfine potential of the pentaquark configuration is
sufficiently attractive compared to that from the sum of a baryon
and a meson. The 4 by 4 matrix form of the expectation value of
the hyperfine factor of the pentaquark configuration generated by
the isospin $\otimes$ color $\otimes$ spin states in
Eq.~(\ref{eq-ICS1-I-1/2}) is given as follows:

\begin{widetext}\allowdisplaybreaks
\begin{align}
&-\langle {\sum}_{i<j}^{5}\frac{1}{m_im_j}\lambda_i^c\lambda_j^c{\sigma}_i\cdot{\sigma}_j \rangle= \nonumber \\
&\left(\begin{array}{cccc}
-\frac{7}{3{m_1}^2}+\frac{1}{2{m_2}^2} +\frac{19}{6m_1m_2}
&  -\frac{\sqrt{2}}{3{m_1}^2}+ \frac{7}{3\sqrt{2}{m_2}^2}-\frac{5\sqrt{2}}{6m_1m_2}
&  \frac{5}{\sqrt{3}{m_1}^2}-\frac{5}{2\sqrt{3}{m_2}^2}-\frac{5}{2\sqrt{3}m_1m_2}
& \frac{\sqrt{5}}{3\sqrt{2}{m_2}^2}+\frac{23\sqrt{5}}{3\sqrt{2}m_1m_2}                             \\
  -\frac{\sqrt{2}}{3{m_1}^2}+ \frac{7}{3\sqrt{2}{m_2}^2}-\frac{5\sqrt{2}}{6m_1m_2}
  & -\frac{8}{3{m_1}^2}+\frac{5}{3{m_2}^2} +\frac{7}{3m_1m_2}
 &  \frac{5\sqrt{2}}{\sqrt{3}{m_1}^2}- \frac{5}{\sqrt{6}{m_2}^2}-\frac{5}{\sqrt{6}m_1m_2}
 & \frac{\sqrt{5}}{3{m_2}^2} - \frac{\sqrt{5}}{3m_1m_2}                                                  \\
   \frac{5}{\sqrt{3}{m_1}^2}-\frac{5}{2\sqrt{3}{m_2}^2}-\frac{5}{2\sqrt{3}m_1m_2}
 &  \frac{5\sqrt{2}}{\sqrt{3}{m_1}^2}- \frac{5}{\sqrt{6}{m_2}^2}-\frac{5}{\sqrt{6}m_1m_2}
  & -\frac{3}{{m_1}^2}+\frac{17}{6{m_2}^2} -\frac{13}{2m_1m_2}
   &   \frac{\sqrt{5}}{\sqrt{6}{m_2}^2} - \frac{\sqrt{5}}{\sqrt{6}m_1m_2}                           \\
\frac{\sqrt{5}}{3\sqrt{2}{m_2}^2}+\frac{23\sqrt{5}}{3\sqrt{2}m_1m_2}
 &   \frac{\sqrt{5}}{3{m_2}^2} - \frac{\sqrt{5}}{3m_1m_2}
 &  \frac{\sqrt{5}}{\sqrt{6}{m_2}^2} - \frac{\sqrt{5}}{\sqrt{6}m_1m_2}
 &  \frac{2}{{m_1}^2}+\frac{1}{{m_2}^2} -\frac{3}{m_1m_2}
 \end{array} \right).
\label{eq-hyper}
\end{align}
\end{widetext}

To compare the expectation values of the hyperfine factor of the
pentaquark with the corresponding sum of a baryon and a meson, we
need to diagonalize -$\langle
{\sum}_{i<j}^{5}\frac{1}{m_im_j}\lambda_i^c\lambda_j^c{\sigma}_i\cdot{\sigma}_j
\rangle $ in Eq.~(\ref{eq-hyper}) and compare it to the possible
decay channels. The diagonalized form of the matrix
-$\langle{\sum}_{i<j}^{5}\frac{1}{m_im_j}\lambda_i^c\lambda_j^c{\sigma}_i\cdot{\sigma}_j
\rangle $ in Eq.~(\ref{eq-hyper}) can be represented as
combinations of terms proportional to $1/{m_1}^2$, $1/{m_2}^2$,
and $1/(m_1m_2)$, respectively. When the fitting mass $m_u$ and
$m_c$ in Table~\ref{fitting-parameter} are used, the ground state
is given as
\begin{align}
-\frac{7.88}{{m_1}^2}+\frac{5.29}{{m_2}^2} -\frac{1.41}{m_1m_2}=-87.3~\rm{(GeV)^{-2}}.
\label{eq-digonal-hyper}
\end{align}

As can be seen in Table~\ref{decay}, the ground state of the
diagonalized hyperfine factor of the pentaquark in
Eq.~(\ref{eq-digonal-hyper}) is slightly more attractive than the
most attractive $p+J/\psi$ decay channel. This attraction is
coming from the term proprotional to $1/m_1m_2$, which originates
from the additional attraction coming from bringing the color
octet component of $p$ and $J/\psi$ together, as noted recently in
Ref.~\cite{Takeuchi:2016ejt}. However, as we will show below, the
attraction is very small and will not compensate for the
additional kinetic energy term that arises from making the
pentaquark state compact compared to the isolated meson baryon
states.

To investigate the mass and the property of the pentaquark with
the variational method, we calculate the Schr\"odinger equation
$H\vert {\Psi}_{\alpha}\rangle = E_{\alpha}\vert
{\Psi}_{\alpha}\rangle$ and diagonalize the $4 \times 4$ matrix.
We find the ground state to be 4087.6 $\rm{MeV}$, which is the sum
of the mass of the $p$ and $J/\psi$ in our model. The wave
function is given as
\begin{align}
\vert {\Psi}_{g}\rangle=&-0.4082\vert R \rangle \vert [I^{\frac{1}{2}}CS]_1 \rangle-0.5773\vert R \rangle \vert [I^{\frac{1}{2}}CS]_2 \rangle \nonumber \\
&+0.7071\vert R \rangle \vert [I^{\frac{1}{2}}CS]_3 \rangle,
\label{eq-ground-I=1/2}
\end{align}
where the variational parameters are given as
$a_1=3.4~\rm{fm^{-2}}$, $a_2=1.4 ~\rm{fm^{-2}}$,
$a_3=11~\rm{fm^{-2}}$ and $a_4 \sim 0$. The first two parameters
and the third parameter correspond to those of the baryon and
meson, respectively, while the last shows that the distance
between the center of mass of the baryon and the meson approaches
infinity. In fact, as we can see from the transformation matrix in
Eq.~(\ref{eq-transform-ICS1}), the ground state, $\vert
{\Psi}_{g}\rangle$, for $I=1/2$ is exactly equal to
-$(p)_1\otimes(J/\psi)_1$ corresponding to $\vert \psi_1\rangle$
in Eq.~(\ref{eq-ICS2-I-1/2}), which means that the ground state
corresponds to the isolated $p$ and $J/\psi$ state in the relative
S-wave.

It is useful to inspect the expectation value of the Hamiltonian
for the state $\vert \psi_1\rangle$ to understand why the
separated $p$ and $J/\psi$ configuration becomes the ground state.
First, the hyperfine potential -$\langle
{\sum}_{i<j}^{5}\frac{1}{m_im_j}\lambda_i^c\lambda_j^c{\sigma}_i\cdot{\sigma}_j
\rangle$ is -$\frac{8}{{m_1}^2}$+$\frac{16}{3{m_2}^2}$, which is
exactly equal to the sum of the expectation value of the $p$ and
$J/\psi$ with the first term (the second) coming from the $p$
($J/\psi$).  Moreover, as discussed before, the lowest eigenvalue
of the hyperfine matrix is not so different from this value,
suggesting that the attraction in the color octet $p$ and $J/\psi$
is not so strong attraction.   As for the confinement potential,
as can be seen from Eq.~(\ref{eq-lambda1})-(\ref{eq-lambda2}) in
the Appendix, the first diagonal components consist of the terms
corresponding to the $p$ and $J/\psi$ only.   Therefore, the only
mass difference between the pentaquark and the $p$ + $J/\psi$
channel comes from the additional kinetic term, which vanishes for
the separated $p+J/\psi$ state. Using the last term in
Eq.~(\ref{eq-kinetic}), one can estimate the additional kinetic
energy to bring the $p$ and $J/\psi$ together.  Taking $a_4 \sim 2
~\rm{fm}^{-2}$, which corresponds to a separation of about  0.7
fm, one obtains an extra kinetic energy of 200 MeV, making the
energy of the compact pentaquark state to be around 4290 MeV. Even
if we allow the other three states to mix, which could bring in
small additional hyperfine attraction, the additional confining
potential will conspire to keep the $(p)_1\otimes(J/\psi)_1$ state
the dominant compact configuration.
 Obviously, such a compact state would just fall apart into the $p+J/\psi$ state and thus not be stable unless the spatial wave function has a small overlap with the final state $p+J/\psi$ \cite{Melikhov:2006ec}.

\begin{table}[htp]
\caption{ The sum of the expectation value of the hyperfine factor
of both a baryon and a meson for the possible decay channel with
respect to $I=1/2$. The third column shows the value for the
fitting mass $m_u$ and  $m_c$.  (unit:$\rm{(GeV)^{-2}}$) }
\begin{center}
\begin{tabular}{c|c|c}
\hline \hline
Decay channel &-$\langle{\sum}_{i<j}^{N}\frac{1}{m_im_j}\lambda_i^c\lambda_j^c{\sigma}_i\cdot{\sigma}_j\rangle$
& Value  \\
\hline
  $pJ/\psi$ & -$\frac{8}{{m_1}^2}$+$\frac{16}{3{m_2}^2}$    &    -86.2   \\
\hline
  ${\Lambda}_cD^*$   &  -$\frac{8}{{m_1}^2}$+$\frac{16}{3m_1m_2}$   & -78.3   \\
 \hline
  ${\Sigma}_c^*D$   &  $\frac{8}{3{m_1}^2}$-$\frac{32}{3m_1m_2}$   & 10.5  \\
 \hline
 ${\Sigma}_cD^*$   & $\frac{8}{3{m_1}^2}$-$\frac{16}{3m_1m_2}$    &  19.8 \\
\hline
${\Sigma}_c^*D^*$   & $\frac{8}{3{m_1}^2}$+$\frac{32}{3m_1m_2}$    &  47.9  \\
\hline \hline
\end{tabular}
\end{center}
\label{decay}
\end{table}

\begin{table}[htp]
\caption{The mass of the excited state of the pentaquark with
$I=1/2$ obtained from the variational method, by diagonalizing the
matrix element of the Hamiltonian in terms of $\vert R \rangle
\vert  \psi_2 \rangle$,  $\vert R \rangle \vert  \psi_3 \rangle$,
and $\vert R \rangle \vert   \psi_4 \rangle$.  ${\Delta}_B$
indicate the binding energy.  The units for the energy and variational parameter are $\rm{GeV}$ and fm$^{-2}$, respectively.  }
\begin{center}
\begin{tabular}{c|c|c|c|c|c}
\hline \hline
 I=1/2 & \multicolumn{5}{|c}{$q^3c\bar{c}$}                                            \\
\hline
 Mass & \multicolumn{5}{|c}{ 4.626}                                            \\
Variational parameters & \multicolumn{5}{|c}{ $a_1$=2.3, $a_2$=1.4,
 $a_3$=4, $a_2$=3.4 }                      \\
\hline
  Decay channel & $pJ/\psi$  & ${\Lambda}_cD^*$   &  ${\Sigma}_c^*D$   &  ${\Sigma}_cD^*$   &                       ${\Sigma}_c^*D^*$                            \\
\hline
 Threshold & 4.088  & 4.298    &  4.408  & 4.471     &     4.548                        \\
\hline
 ${\Delta}_B$ & 0.538 & 0.328   &  0.218  &  0.155   &  0.078                        \\
\hline \hline
\end{tabular}
\end{center}
\label{mass}
\end{table}

As any configuration generated with $\vert \psi_1\rangle$ is
dominated by the fall apart $p+J/\psi$ state, we need to
investigate whether the excited state can be compact and
quasi-stable. To accomplish this, we consider the $\vert
\psi_2\rangle$, $\vert \psi_3\rangle$, and $\vert \psi_4\rangle$
in Eq.~(\ref{eq-ICS2-I-1/2}) without $\vert \psi_1\rangle$. The
detailed property of the excited state of this state is given in
Table~\ref{mass}. Due to the quantum numbers, except for the
$p+J/\psi$ configuration, the excited states can not be written as
a sum of a single baryon and meson state. Hence, we find a 
compact state. However, it can decay into several baryon and meson
decay channels and is not stable. As for the color spin part of
the potential
-$\langle{\sum}_{i<j}^{5}\frac{1}{m_im_j}\lambda_i^c\lambda_j^c{\sigma}_i\cdot{\sigma}_j
\rangle $, we find that this state has the following form;
\begin{align}
-\frac{1.27}{{m_1}^2}-\frac{0.45}{{m_2}^2} -\frac{5.38}{m_1m_2}=-23.4~\rm{(GeV)^{-2}}.
\label{eq-excited-digonal-hyper}
\end{align}
While the diagonalized hyperfine factor are less attractive than
that of the $p+J/\psi$ and ${\Lambda}_c+D^*$ decay channels, it is
still more attractive than other decay channels. Nevertheless, the
reason why the excited state has energy larger than any decay
channel is due to the large contribution from the confining
potential. As discussed in the Appendix, the sum of the color
matrix are all equal for the four orthonormal states. However, due
to the interplay with the kinetic term, the confining part of the
potential is most attractive in the $p+J/\psi$ channel. The
contributions from the kinetic, confinement and hyperfine
interaction terms for the excited pentaquark state as well as
separated baryon meson states are summarized in
Table\ref{energy-contribution}. The large confinement contribution
for the pentaquark state  can be seen in the
Table\ref{energy-contribution}. The obtained mass is too large for
it to be the one of the recently observed pentaquark states.
Moreover, it will decay to all possible baryon meson state and not
be stable.

\begin{table}[htp]
\caption{The values of each energy term of the excited state of
the pentaquark and the sum of a baryon and a meson in decay
channel. $\Delta$E is the difference between the pentaquark and
its decay channel in each term. (unit:$\rm{Mev}$) }
\begin{center}
\begin{tabular}{c|c|c|c|c}
\hline \hline
 Pentaquark            & Kinetic    & Comfinement & Hyperfine  & Sum \\
\hline
The excited state      & 1144.3 & 1238 & -52.1   & \\
\hline
 Decay channel        & Kinetic    & Comfinement & Hyperfine & Sum  \\
\hline
 $pJ/\psi$              &   1190.5 & 745.8 & -145.1 &  \\
 $\Delta$E              &  -46.2    &  492.2  &  93  &      \\
\hline
 ${\Lambda}_cD^*$   &  1192.7 & 982.2 & -173.1 & \\
  $\Delta$E              & -48.4    & 255.8  &  121 &       \\
\hline
 ${\Sigma}_c^*D$     & 1105.3 & 1055.1  & -48.6  &  \\
  $\Delta$E              &  39    &   182.9   & -3.5  &       \\
\hline
 ${\Sigma}_cD^*$     & 1046.5 & 1102.9  &25.8  &  \\
  $\Delta$E             & 97.8     & 135.1     &  -77.9 &        \\
\hline
 ${\Sigma}_c^*D^*$  & 993.1   & 1157    & 101.4  & \\
 $\Delta$E              &  151.2    &  81      & -153.5 &        \\
 \hline \hline
\end{tabular}
\end{center}
\label{energy-contribution}
\end{table}

\section{Summary}

To understand the possible quark configuration of the recently
observed hidden charm pentaquark state, we systematically
construct the isospin $\otimes$ color $\otimes$ spin pentaquark
states containing two heavy quark and antiquark with $I=1/2$ and
$S=3/2$ that satisfy the Pauli principle. We systematically derive
the isospin $\otimes$ color $\otimes$ spin states from the color
and spin coupling scheme, which is based on the permutation group
property. We found that there are four orthonormal state, one of
which is the color, spin and isospin corresponding to the proton
and $J/\psi$. Then, by using a spatial trial wave function that is
suitable for describing the decay into a baryon and meson state,
we perform the variational method to obtain the lowest mass state
of the pentaquark with $I=1/2$ and $S=3/2$. We found that the
ground state is the isolated $p+J/\psi$ state and that any compact
configuration will also be dominated by the same baryon and meson
state, which will thus fall apart decay to the ground state. We
further calculate the mass with a excited state, involving the
other isospin $\otimes$ color $\otimes$ spin states which are
orthonormal to the ground state. The mass of the compact exited
state is found to be well above all baryon meson decay channel and
not stable. We are therefore led to conclude that the recently
observed pentaquark state can not be a compact multiquark state
within the conventional constituent quark model with only
confining and color spin interaction. There could still be
intrinsic three or four body quark interaction that might change
the situation. Also, hadronic molecular configurations originating
from meson exchange can certainly not be handled in the present
picture. All such works are topics for future works.

\section{Acknowledgments}

This work was supported by the Korea National Research Foundation
under the grant number KRF-2011-0020333 and KRF-2011-0030621. The
work of W.S. Park was supported in part by the Yonsei University
Research Fund (Post Doc. Researcher Supporting Program) of 2014
(project no.: 2014-12-0139). The work of S. Cho and W. S. Park was
supported by the National Research Foundation of Korea (NRF) grant
funded by the Korea government (MSIP) (No. 2016R1C1B1016270).

\appendix

\section{}

In this Appendix, we will present the matrix element of
$\lambda_i^c\lambda_j^c$  ($i$$<$$j$=1$\sim$5) of the pentaquark
in terms of a four dimensional matrix generated by the states
$\vert \psi_1\rangle$,  $\vert \psi_2\rangle$, $\vert
\psi_3\rangle$, and $\vert \psi_4\rangle$ in
Eq.~(\ref{eq-ICS2-I-1/2}).

 a) ($i$,$j$)=(1,2), (1,3), or (2,3);
\begin{align}
\langle \lambda_i^c\lambda_j^c \rangle=\left(\begin{array}{cccc}
-\frac{8}{3}       &    0     & 0           & 0                                     \\
0                  &   -\frac{2}{3}            & 0              & 0                  \\
0                  &            0              & -\frac{2}{3}       &      0         \\
0                  &             0         &  0         & -\frac{2}{3}               \\
          \end{array} \right),
\label{eq-lambda1}
\end{align}

b) ($i$,$j$)=(1,4), (1,5), (2,4), (2,5), (3,4), or (3,5);
\begin{align}
\langle \lambda_i^c\lambda_j^c \rangle=\left(\begin{array}{cccc}
0       &    0     & 0           & 0                                     \\
0                  &   -2          & 0              & 0                  \\
0                  &            0              & -2       &      0         \\
0                  &             0         &  0         & -2             \\
          \end{array} \right),
\label{eq-lambda2}
\end{align}

 c) ($i$,$j$)=(4,5);
\begin{align}
\langle \lambda_i^c\lambda_j^c \rangle=\left(\begin{array}{cccc}
-\frac{16}{3}       &    0     & 0           & 0                                     \\
0                  &   \frac{2}{3}            & 0              & 0                  \\
0                  &            0              & \frac{2}{3}       &      0         \\
0                  &             0         &  0         & \frac{2}{3}               \\
          \end{array} \right).
\label{eq-lambda3}
\end{align}
It is easily seen that
$\langle{\sum}_{i<j}^{5}\lambda_i^c\lambda_j^c \rangle $=
-40/3$I$, where the $I$ is identity matrix.

In the case of a baryon, $\langle {\sum}_{i<j}^{3}
\lambda_i^c\lambda_j^c \rangle =-8$ coming from the color singlet
state $\frac{1}{\sqrt{6}}\epsilon_{ijk}q^i(1)q^j(2)q^k(3))$. For a
meson state, $\langle \lambda_4^c\lambda_5^c \rangle= -16/3$ with
the color state $\bar{q}_i(4)q^i(5)$. These values are the first
diagonal components in the above matrix elements. Hence, as
pointed out before, we find that the first diagonal term of
$\langle {\sum}_{i<j}^{5} \lambda_i^c\lambda_j^c \rangle$ of the
pentaquark is just the sum of those of the baryon and meson. In
fact, as far as this color matrix is concerned, all the four sum
of diagonal matrix elements have the same value. However,
depending on the spatial wave function, the matrices for the
confining potential will have different weighting factors coming
from spatial wave functions and their sum will no longer be
proportional to the identity matrix. If the kinetic terms are
considered, it is energetically more favorable to maximize the
attraction in the $p$ and $J/\psi$ channel, which makes it the
most attractive state even for compact configurations.

\end{document}